\documentclass[aps,prl,twocolumn,nofootinbib]{revtex4}
\usepackage{epsfig}
\usepackage{graphicx}
\usepackage{amsmath}
\usepackage{amsfonts}
\usepackage{amssymb}
\usepackage{bm}
\begin{document}
\title{On the Stacking Charge Order in $\rm NaV_2O_5$}
\author{Gennady Y. Chitov}
\altaffiliation{Address after August 1, 2004:
Department of Physics and Astronomy, Laurentian University,
Sudbury, Ontario P3E 2C6, Canada}
\affiliation{Department 7.1-Theoretical Physics,
University of Saarland, Saarbr\"ucken D-66041, Germany}
\author{Claudius Gros}
\affiliation{Department 7.1-Theoretical Physics,
University of Saarland, Saarbr\"ucken D-66041, Germany}

\date{\today}

\begin{abstract}
We propose a mechanism for the observed stacking charge order
in the quarter-filled ladder compound $\rm NaV_2O_5$.
Via a standard mapping of the charge degrees of freedom onto
Ising spins we explain the stacking order as a result of competition
between couplings of the nearest and next-nearest planes with the
4-fold degenerate super-antiferroelectric in-plane order.
\end{abstract}

\maketitle
\begin{figure}
\epsfig{file=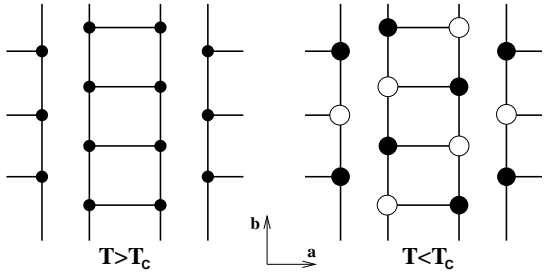,width=0.40\textwidth,angle=0}
\caption{$\rm NaV_2O_5$: ladders in the $ab$-plane. In the disordered
phase each electron is equally distributed between left/right sites on a given
rung (left panel), while below $T_c$ electrons (filled black circles)
order as shown in the right panel.
}
\label{NaVO}
\end{figure}

There has been a great interest in recent years from both theorists and
experimentalists in the insulating transition-metal compound
$\rm NaV_2O_5$ \cite{Lem03}. This material provides a rather unique
example of a correlated electron system, where the interplay of charge
and spin degrees of freedom results in a phase transition into a phase
with coexistent spin gap and charge order. $\rm NaV_2O_5$ is the only
known so far quarter-filled ladder compound \cite{Smo98}. 
Each individual rung of the ladder is occupied
by single electron which is equally distributed between its left/right
sites in the disordered phase (see Fig.\ \ref{NaVO}). At $T_c=34\,{\rm K}$
this compound undergoes a phase transition when a spin gap opens,
accompanied by charge ordering \cite{Iso96,Fuj97,Sma02,Grenier02}.
The experimentally observed two-dimensional (2D) long-range charge order
in $\rm NaV_2O_5$ \cite{Sma02,Grenier02} (the $ab$-plane order is shown in
Fig.\ \ref{NaVO}) is {\it super-antiferroelectric} (SAF),
as we have pointed out recently \cite{CGIs}.
The theory of spin-SAF transition put forward by us
\cite{CGbig03,CGIs,ExactDia} is adequate in accounting for,
even quantitatively, various aspects of the transition in $\rm NaV_2O_5$.
It deals, however, with a single ($ab$) plane, leaving aside the question
of charge ordering along the third ($c$) direction. The phase transition
in $\rm NaV_2O_5$ quadruples the unit cell in $c$-direction (the supercell
of the ordered phase is \textrm{$2a \times 2b \times 4c$}), and the
recent X-ray experiments \cite{Sma02,Grenier02} revealed peculiar
ordering patterns in $c$-direction (stacking order) of the
super-antiferroelectrically charge-ordered planes. Here we present a
model which provides the explanation for the observed stacking order in
$\rm NaV_2O_5$.

An insulating quarter-filled ladder system, with
electrons localized on the rungs of the ladder, 
can be mapped on an effective spin-pseudospin model, where the Ising
pseudospin ($\mathcal T^x$) represents left/right positions
of the charge on a given rung, similar to the standard
pseudospin approach to the order-disorder-type phase transitions
\cite{Blinc74}. The 2D effective spin-pseudospin Hamiltonian,
able to describe the spin-SAF transition in $\rm NaV_2O_5$, was given
in \cite{CGIs}. Since the present work is concerned with the physics of
the charge order, we will discuss here exclusively the Ising sector of
the full spin-pseudospin model.

\begin{figure}
\epsfig{file=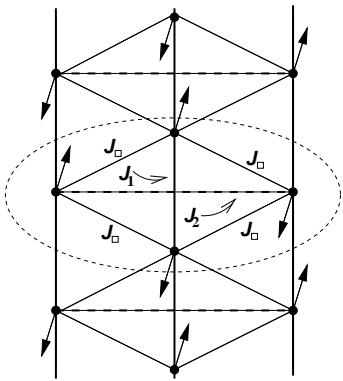,width=0.25\textwidth,angle=0}
\caption{$ab$-plane of coupled ladders mapped onto an effective 
2D lattice. A vertical line and a dot represent a single ladder and its rung.
The Ising pseudospin (up/down) represents the position (left/right) of the
electron on a given rung.
}
\label{EffLat}
\end{figure}

\begin{figure}
\epsfig{file=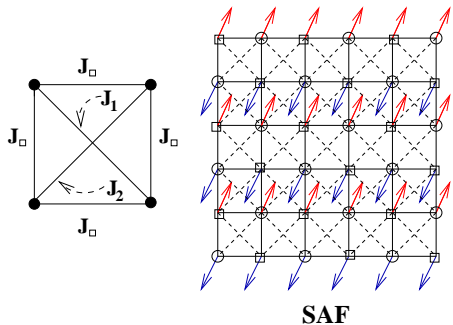,width=0.33\textwidth,angle=0}
\caption{Couplings on an elementary plaquette of the 2D nn and nnn
Ising model (left) and an example of the SAF order (right). An
plaquette on the square lattice corresponds to the encircled region
in Fig.\ \ref{EffLat}.
}
\label{PlaqSAF}
\end{figure}

Within the spin-pseudospin formalism, the $ab$-plane of
coupled ladders can be mapped on the effective lattice shown in
Fig.\ \ref{EffLat}, which in its turn we smoothly map onto a
more conventional square lattice.
It is then easy to identify the charge order
in the $ab$-plane (see Figs.\ \ref{NaVO} and \ref{COex} below)
as the SAF phase \cite{FanWu69} of the 2D nearest neighbor (nn) and
next-nearest neighbor (nnn) Ising model, shown in Fig.\ \ref{PlaqSAF}.
Since the Ising couplings
$J_{\sharp}=J_{\text{\tiny $\square$}},J_1,J_2$ 
originate from the Coulomb repulsion, 
we assume them to be antiferro (AF), i.e., $J_{\sharp}>0$. 
SAF is the ground state of the 2D nn and nnn Ising model,
if \cite{CGIs}
\begin{equation}
\label{SAFGS}
J_1+J_2 >|J_{\text{\tiny $\square$}}|,\quad\textrm{and}\quad J_{1,2}>0
\end{equation}
The SAF state can be viewed as two superimposed
antiferromagnetically ordered sublattices (circled/squared sites
shown in the right panel of Fig.\ \ref{PlaqSAF}), and it is 4-fold
degenerate, since each of these sublattices can be flipped independently.

Some possible patterns of the stacking charge order in $\rm NaV_2O_5$,
determined from the X-ray experiments \cite{Grenier02}, are shown in
Fig.\ \ref{COex}. In terms of the effective Ising model this translates
into the 3D pseudospin ordering patterns depicted in Fig.\ \ref{AxSAF}.
To explain the mechanism of these types of order, let us consider the
``minimal'' 3D nn and nnn Ising model with the Hamiltonian
\begin{equation}
\label{NNNIs}
H =
\frac12 \sum_{\mathbf{k},\mathbf{l} }
J_{\mathbf{k} \mathbf{l}}
\mathcal T^x_{\mathbf{k}} \mathcal T^x_{\mathbf{l}}
\end{equation}
where the bold variables denote lattice vectors. The sum includes spins
on the nn sites coupled via $J_{\text{\tiny $\square$}}, J_3$ and on the
nnn sites coupled via $J_{1,2,4}$ (cf. Figs.\ \ref{PlaqSAF},\ref{AxSAF}).
This model reduces to a more familiar 3D ANNNI model \cite{Liebmann86}
when the diagonal couplings $J_{1,2}$ are absent. We will be interested
in the case of the plaquette couplings satisfying (\ref{SAFGS}), i.e.,
when the planes are SAF-ordered in the ground state, and a
frustrating nnn-interplane coupling $J_4>0$.

Note, that the four patterns \textbf{A}-\textbf{D} of the SAF state
can be divided into two pairs of the ``AF counterparts''
(\textbf{A},\textbf{B}) and (\textbf{C},\textbf{D}) in the sense that in
the presence of an AF coupling between two SAF-ordered planes,
the counterparts \textbf{A} and \textbf{B} (or \textbf{C} and \textbf{D})
minimize the energy without inter-plane frustrations
(cf. Fig.\ \ref{AxSAF}).

From energy considerations one finds that, if $J_4>|J_3|/2$, the model has
a 16-fold degenerate ground state composed of the SAF-ordered planes
stacked with a period of four lattice spacings in $z(c)$-direction. We
will denote this 3D order as SAF$\times4$. (Note that the in-plane SAF
phase itself can take either $2 \times 1$ pattern, or  $1 \times 2$.)
The number of frustrated inter-plane bonds ($J_3$) in the SAF$\times4$
phase is 8 per 4 stacked elementary cubes of the lattice (in average
2 $J_3$-bonds per elementary cube). This phase
can be realized via four 4-fold degenerate stacking patterns:
\begin{eqnarray}
\nonumber
&~&\langle \textbf{AABB} \rangle,~~\textrm{type}~\textbf{I} \\ \nonumber
&~&\langle \textbf{ACBD} \rangle,~~\textrm{type}~\textbf{II} \\ \nonumber
&~&\langle \textbf{ADBC} \rangle,~~\textrm{type}~\textbf{III} \\ \nonumber
&~&\langle \textbf{CCDD} \rangle,~~\textrm{type}~\textbf{IV}. \nonumber
\end{eqnarray}%
4-fold degeneracy of each of these stacking patterns comes from
translations along the stacking direction, or,
in terms of the above notations, from cyclic permutations inside
the angular brackets. So, the two particular realizations
of the SAF$\times4$ state shown in the left and right panels of
Fig.\ \ref{AxSAF} belong to the patterns \textbf{II} and \textbf{IV},
respectively.
\begin{figure}
\epsfig{file=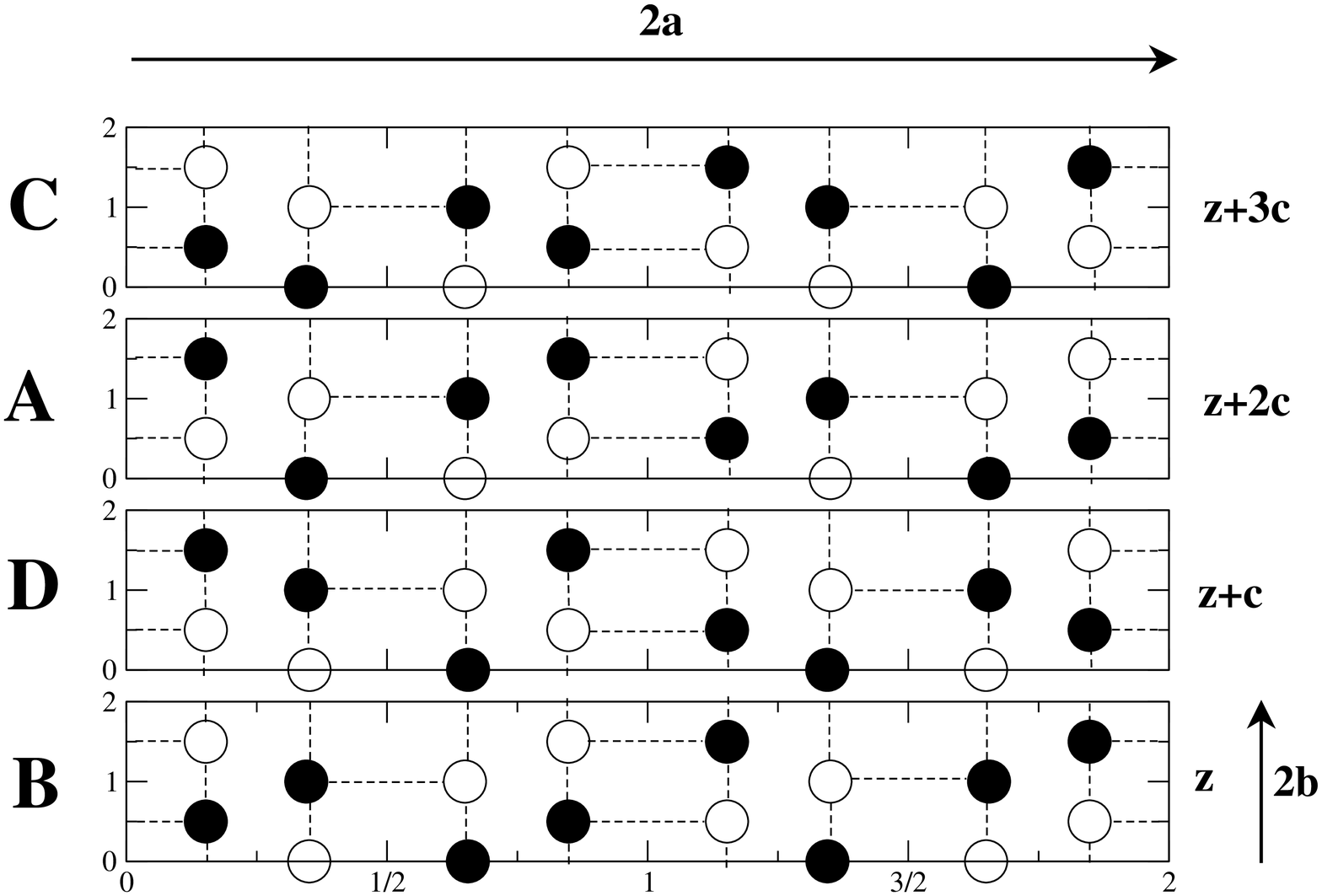,width=0.32\textwidth,angle=0}
\\[-.5cm]
\epsfig{file=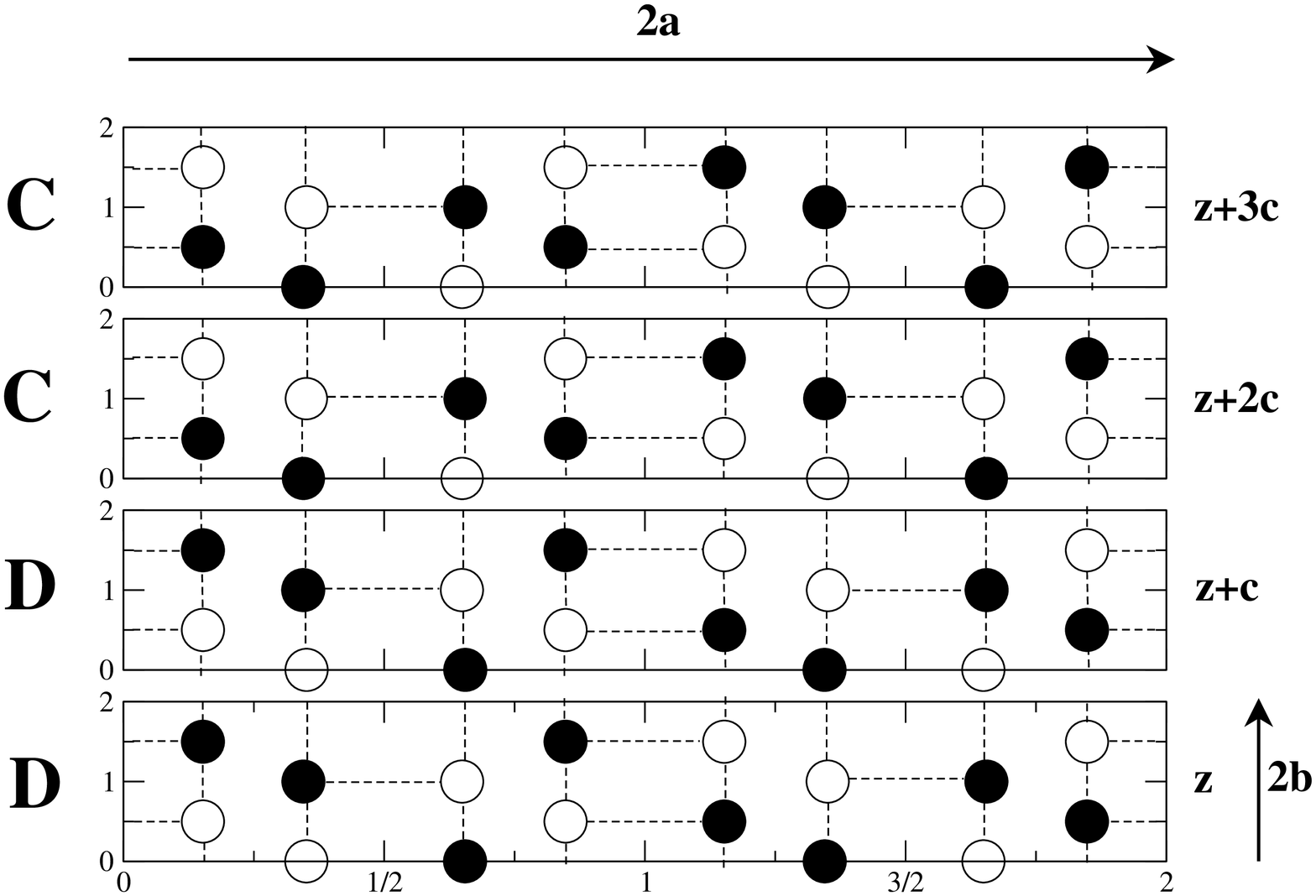,width=0.32\textwidth,angle=0}
\\[-.5cm]
\caption{Possible patterns of the stacking charge order in
$\rm NaV_2O_5$ from the X-ray experiments
(reproduced from Ref.\cite{Grenier02}): $\langle \textbf{BDAC}\rangle$
(and $\langle \textbf{DBCA}\rangle$, not shown) [top];
$\langle\textbf{DDCC} \rangle$ (and $\langle \textbf{BBAA}\rangle$,
not shown) [bottom]. Each layer (in $ab$-plane) labeled by
\textbf{A}-\textbf{D} corresponds to a particular realization of the
4-fold degenerate SAF state.
}
\label{COex}
\end{figure}

If $J_4<|J_3|/2$ the model can have two possible 4-fold degenerate ground
states SAF$\times1$ or SAF$\times2$, depending on the sign of $J_3$.
If the nn-interplane coupling is AF $J_3>0$, the model's ground state
SAF$\times 2$ can be realized via two 2-fold degenerate stacking patterns
$\langle \textbf{AB} \rangle$ or $\langle \textbf{CD} \rangle$. 
For a ferromagnetic nn-interplane coupling
$J_3<0$, the 4-fold degenerate ground
state SAF$\times 1$ is simply one of the four possible SAF
patterns stacked in $c$-direction. In each of the phases SAF$\times1$ or
SAF$\times2$ there are 4 frustrated inter-plane bonds ($J_4$) per
elementary cube of the lattice. Thus, to summarize the ground-state
phases:
\begin{eqnarray}
\label{SAF4}
&J_4&>|J_3|/2:~~\textrm{SAF$\times4$} \\
\label{SAF2}
&J_4&<J_3/2:~~~~\textrm{SAF$\times2$} \\
\label{SAF1}
&J_4&<-J_3/2:~~\textrm{SAF$\times1$}
\end{eqnarray}
\begin{figure}
\epsfig{file=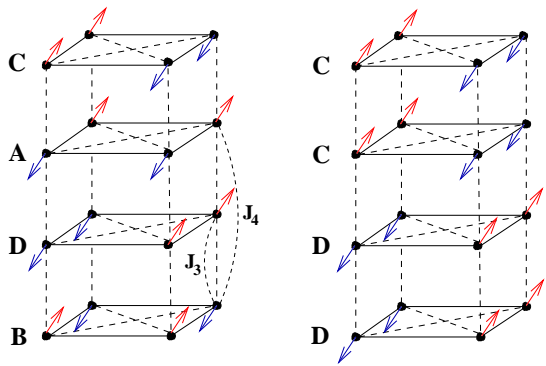, width=0.40\textwidth,angle=0}
\caption{Two particular realizations of the 16-fold degenerate
ground state of the 3D nn and nnn Ising model
(\ref{NNNIs}) at $J_4>|J_3|/2$. The depicted Ising orders correspond
to the charge ordering patterns shown in Fig.\ \ref{COex}.
}
\label{AxSAF}
\end{figure}

The model (\ref{NNNIs},\ref{SAF4})  provides the explanation for the charge
ordering in $\rm NaV_2O_5$ found in the most recent X-ray studies
\cite{Grenier02}, carried out at ambient pressure deep in the ordered
phase (at $T=13\,{\rm K}$). Those authors found that the sum of patterns
$\langle \textbf{BDAC}\rangle~+~\langle \textbf{DBCA}\rangle$
(i.e.,  $\langle\textbf{II}+\textbf{III}\rangle$ type)
and/or of $\langle\textbf{DDCC} \rangle~+~\langle \textbf{BBAA}\rangle$
(i.e.,  $\langle\textbf{IV}+\textbf{I}\rangle$ type),
makes the best fit to the scattering data, and not a single pattern.
This implies that the actual stacking charge order in $\rm NaV_2O_5$
accommodates all those (degenerate) patterns with stacking faults
\cite{Sma02,Grenier02}.

An analysis of stacking faults between the
various (totally 16) faultless patterns shows that they cost 
different energy. For instance, if $J_3<0$ the
least energetically expensive (the energy is $(J_4-|J_3|/2)N^2$ with
respect to the ground state for $N\times N\times N$ lattice)
are faults of the type
$\langle\textbf{II}~\textrm{or}~\textbf{III}\rangle \bullet
\langle\textbf{I}~\textrm{or}~\textbf{IV}\rangle$, where $\bullet$
indicates the position of the fault. The examples are:
\begin{eqnarray}
\label{F1}
&~&....\textbf{ACBD} \bullet \textbf{DCCD}....~~  \\ \nonumber
&~&....\textbf{ADBC} \bullet \textbf{AABB}....
\end{eqnarray}%
The faults of the type $\langle\textbf{II}~\textrm{or}~\textbf{III}\rangle \bullet
\langle\textbf{II}~\textrm{or}~\textbf{III}\rangle$,
$\langle\textbf{I}\rangle \bullet \langle\textbf{I}\rangle$,
$\langle\textbf{IV}\rangle \bullet \langle\textbf{IV}\rangle$, like, e.g.,
\begin{eqnarray}
\label{F2}
&~&....\textbf{ACBD} \bullet \textbf{DACB}....~~  \\ \nonumber
&~&....\textbf{ACBD} \bullet \textbf{DBCA}....\\ \nonumber
&~&....\textbf{AABB} \bullet \textbf{BAAB}....
\end{eqnarray}%
cost twice more energy than those of the type (\ref{F1}). There are of
course many other possible types of stacking faults (with higher energies)
which we will not discuss. From energy consideration, we can conclude
that at low temperature faults of the type (\ref{F1}) prevail.
This result together with the experimental findings \cite{Grenier02}
lead us to the conclusion that in $\rm NaV_2O_5$ all
four possible types of ordering (\textbf{I-IV}) occur indiscriminately,
with minimal-energy stacking faults, e.g.,
$\langle\textbf{II}\rangle \bullet \langle\textbf{I}\rangle \bullet
\langle\textbf{III}\rangle \bullet \langle\textbf{IV}\rangle$.

Above we assumed $J_3<0$, which might appear odd, since all Ising couplings
originate from the Coulomb repulsion, i.e., they are antiferro. In fact such
ferromagnetic coupling is an effective coupling replacing interactions between
several charges in order to keep the model ``minimal''\footnote{
\label{decoup} For single layer a ``truly minimal'' model, still having
the SAF phase, would be that with $J_{\text{\tiny $\square$}}=0$. In this
case it consists of two decoupled superimposed (AF) Ising lattices. Then
our model (\ref{NNNIs}) with nn and nnn interactions between layers
reduces to \textit{two identical} inter-penetrating decoupled 3D ANNNI
models, cf. Figs.\ \ref{PlaqSAF},\ref{AxSAF}. Let us call this limit the
A$\otimes$A model for brevity.}
In this sense $J_4$ is an effective coupling as well, both $J_3$ and $J_4$
should be viewed as phenomenological parameters whose signs and relative
strengths are chosen to agree with experiments. We can however suggest a
simple mechanism resulting in an effective $J_3<0$. Let us
consider two nn ladders (two chains in terms of the effective lattice
shown in Fig.\ \ref{EffLat}) separated by a lattice unit $c$ in the
stacking direction. On a single plaquette in the $bc$-plane, let us take
into account the nn couplings $J_1$ (along $(0,1,0)$-direction)
and $\tilde J_3$ (along $(0,0,1)$-direction), with
($J_1>J_3$), in addition to a nnn-coupling $J_d$ along
plaquette's diagonals (along $(0,\pm 1,1)$-directions).
A simple analysis shows that two nearest stacked chains 
tend to align ferromagnetically in
$c$-direction when $J_d>\tilde J_3/2$. We can therefore
take $J_3=\tilde J_3-2J_d$ as the
single \textit{effective} nn coupling, as shown in
Fig.\ \ref{AxSAF}. Considering the actual lattice distances
in $\rm NaV_2O_5$ and the distance-dependence of the Coulomb
repulsion one finds $J_3<0$.

Experiments on $\rm NaV_2O_5$ under pressure \cite{Ohwada01} show
that the in-plane SAF charge order is robust and does not change,
while at the pressure $P_c =0.92\,{\rm GPa}$ a transition of the
ground state charge order from SAF$\times4$ into SAF$\times1$ occurs.
Thus, in $\rm NaV_2O_5$ the pressure dependence of in-plane couplings
is ``non-critical'', i.e., condition (\ref{SAFGS}) is satisfied, while
the ratio $\kappa \equiv J_4/J_3$ of the couplings between planes is
more sensitive to pressure.
As follows from (\ref{SAF4},\ref{SAF1}) this
ratio reaches the frustration point $\kappa_c=-1/2$ at $P_c$,
and then the ground state changes.\footnote{\label{J3} In connection to
what was said before about $J_3$, we should point out that experiments
do not rule out $J_3>0$ at ambient pressure since SAF$\times4$
order is insensitive to the sign of $J_3$, as long as (\ref{SAF4})
is satisfied. Then $J_3$ must decrease under pressure such that at $P_c$
it is already ferromagnetic and $\kappa=\kappa_c$. We find
such a possibility rather exotic, and will not consider it here. Note
also that if $J_3>0$ at ambient pressure, the above analysis
of energies of stacking faults should be modified accordingly.}

Another very interesting feature of the charge ordering under pressure
in $\rm NaV_2O_5$ is the existence of a region with numerous
higher-order commensurate superstructures SAF$\times \frac mn$ (where
$m,n$ are integers), i.e., a devil's staircase region above $P_c$ on the
temperature-pressure plane \cite{Ohwada01}. Ohwada and co-workers
noticed resemblance between the experimental phase diagram and that
of the 3D ANNNI model. (For reviews on that model see, e.g.,
Refs.\cite{Liebmann86,Selke88}.)
However an explanation of the charge ordering
in $\rm NaV_2O_5$ in the framework of the ANNNI model is incorrect, as
the latter cannot have the in-plane SAF order in principle. The minimal model
to reproduce the observed experimental results is
what we call the A$\otimes$A model (see footnote \ref{decoup}):
two superimposed square lattices create
the observed overall SAF order in the individual
layers, while the period of the charge order in the stacking direction is 
the same as in a single 3D ANNNI model.

Considering the geometry of the original $\rm NaV_2O_5$ lattice,
one can see that $J_{\text{\tiny $\square$}}$ is indeed rather
small in comparison to $J_1$,\footnote{\label{weakdia}
The same arguments suggest a weak diagonal $J_2$, which is,
however, reinforced effectively due to the bilinear spin-pseudospin
coupling in $\rm NaV_2O_5$ \cite{CGIs}.}
so the minimal A$\otimes$A model appears to be adequate
for the description of the charge ordering in that compound.

The more complicated model (\ref{NNNIs}) with
$J_{\text{\tiny $\square$}} \neq 0$ and (\ref{SAF4}) satisfied,
is expected, from mean-field considerations, to also show
a sequence of commensurate phases originated
from the frustration point $\kappa_c=-1/2$ separating the SAF$\times4$
and SAF$\times1$ ground states, similarly to the 3D ANNNI model.
However to substantiate this suggestion, a separate study of the
temperature phase diagram of (\ref{NNNIs}) is warranted.

So far we have been discussing the charge in $\rm NaV_2O_5$ in terms of
Ising models. The full spin-pseudospin Hamiltonian of the problem is more
involved \cite{CGIs}, since it includes also a transverse field in the
Ising sector plus coupling of the charge (Ising pseudospin) to the spin
degrees of freedom. The present 3D extension of the Ising sector can be
treated along the lines of our earlier analyses \cite{CGbig03,CGIs},
resulting in simultaneous appearance of the charge (Ising) order and
spin gap.

\textit{Conclusions:} We propose a mechanism for the stacking
charge order in $\rm NaV_2O_5$. It is a result of competition
between couplings of the nearest and next-nearest planes with the 4-fold
degenerate SAF in-plane order. The simplest effective model resulting in
the observed charge ordering patterns consists of two decoupled
inter-penetrating 3D ANNNI models (the A$\otimes$A model).
%
%
\begin{acknowledgments}
We thank S. Grenier for providing us with Fig.\ \ref{COex} and
helpful communications.
This work is supported by the German Science Foundation.
\end{acknowledgments}
%



\begin{thebibliography}{}
%
\bibitem{Lem03} For a review, see P. Lemmens, G. G\"untherodt, and C. Gros,
Phys. Reports. {\bf 375}, 1 (2003).
%
\bibitem{Smo98} H. Smolinski, C. Gros, W. Weber, U. Peuchert, G. Roth,
      M. Weiden, C. Geibel, 
Phys. Rev. Lett. {\bf 80}, 5164 (1998).
%
\bibitem{Iso96} M. Isobe and Y. Ueda,
                J. Phys. Soc. Japn. {\bf 65}, 1178 (1996).
%
\bibitem{Fuj97} Y. Fujii, H. Nakao, T. Yosihama, M. Nishi, K. Nakajima,
K. Kakurai, M. Isobe, Y. Ueda, and H. Sawa,
                J. Phys. Soc. Japn. {\bf 66}, 326 (1997).
%
\bibitem{Sma02} S. van Smaalen, P. Daniels, L. Palatinus,
         and R.K. Kremer, Phys. Rev. B \textbf{65}, 060101 (2002).
%
\bibitem{Grenier02} S. Grenier, A. Toader, J. E. Lorenzo, Y. Joly,
B. Grenier, S. Ravy, L. P. Regnault, H. Renevier, J. Y. Henry,
J. Jegoudez, and A. Revcolevschi,
Phys. Rev. B \textbf{65}, 180101(R) (2002).
%
\bibitem{CGIs} G.Y. Chitov and C. Gros, cond-mat/0401295.
%
\bibitem{CGbig03} G.Y. Chitov and C. Gros,
Phys. Rev. B \textbf{69}, 104423 (2004).
%
\bibitem{ExactDia} C. Gros and G.Y. Chitov, cond-mat/0403263.
%
\bibitem{Blinc74} R. Blinc and B. \v{Z}ek\v{s},
\textit{Soft Modes in Ferroelectrics and Antiferroelectrics},
(North-Holland Publishing Co., Amsterdam, 1974).
%
\bibitem{FanWu69} C. Fan and F.Y. Wu,
Phys. Rev. {\bf 179}, 560 (1969).
%
\bibitem{Liebmann86} For a review, see R. Liebmann,
\textit{Statistical Mechanics of Periodic Frustrated Ising Systems}
(Springer, Berlin, 1986).
%
\bibitem{Ohwada01} K. Ohwada, Y. Fujii, N. Takesue, M. Isobe, Y. Ueda,
H. Nakao, Y. Wakabayashi, Y. Murakami, K. Ito, Y. Amemiya, H. Fujihisa,
K. Aoki, T. Shobu, Y. Noda, and N. Ikeda,
Phys. Rev. Lett. \textbf{87}, 086402 (2001).
%
\bibitem{Selke88} W. Selke, Phys. Rep.  {\bf 170}, 213 (1988).
%
\end{thebibliography}
\end{document}